\documentclass[preprint,showpacs,preprintnumbers,amsmath,amssymb]{revtex4}
\usepackage{amsmath,amssymb,graphicx} 
\usepackage{epsf}
\usepackage{epstopdf}
\usepackage {amssymb}
\newcommand{\nc}{\newcommand}
\nc{\ba}{\begin{eqnarray}}
\nc{\ea}{\end{eqnarray}}
\nc{\ga}{\gamma}

\nc{\x}{{\bf{x}}}
\nc{\y}{{\bf{y}}}

\input{epsf.sty}

\begin{document}

\title{ Lensing and CMB Anisotropies by Cosmic Strings at a Junction}

\author{Robert Brandenberger}
\email{rhb@physics.mcgill.ca}
\author{Hassan Firouzjahi }
\email{firouz@physics.mcgill.ca}
\author{Johanna Karouby}
\email{karoubyj@physics.mcgill.ca}
\affiliation{Physics Department, McGill University, 3600 University Street, Montreal, Canada, H3A 2T8 }

\begin{abstract}
The metric around straight arbitrarily-oriented cosmic strings forming a stationary junction is obtained at the linearized level. It is shown that the geometry is flat. The sum rules for lensing by this configuration
and the anisotropies of the CMB  are obtained.

\vspace{0.3cm}

Keywords : Cosmic strings, Cosmology 
\end{abstract}

\maketitle

\section{Introduction}
There has been a renewal of interests in cosmic strings  \cite{Kibble:2004hq}.
This is partially due to the realization that in models of brane inflation cosmic strings, and not monopoles and 
domain walls, are copiously produced \cite{Sarangi:2002yt, Jones:2003da}.
In these models of inflation, the inflaton is the distance between
a D3-brane and an anti D3-brane \cite{Dvali:1998pa} . There is an attractive force between the brane and anti-brane. Inflation ends when they collide and annihilate each other. The end product
is a network of one-dimensional defects in the form of fundamental strings, F-strings, and D1-branes, D-strings. These can further combine to
form a bound states of p F-strings and q D-strings, (p,q)-strings, for
integer p an q.

In more developed models of brane inflation, inflation happens
in a warped region inside the string theory compactification \cite{Kachru:2003sx}. This way, the scale of inflation as well as the effective tension of cosmic strings, $\mu$, are considerably smaller than the string scale \cite{Firouzjahi:2005dh}. One can easily saturate the current bounds
on $G\mu$, the dimensionless number corresponding to the cosmic string 
tension \cite{Wyman:2005tu}.

When two (p,q) cosmic strings intersect generally a junction is formed. This is due to charge conservation. This is in contrast to U(1) gauge cosmic strings. When two U(1) gauge cosmic strings intersect, they usually exchange partners and intercommute with probability close to unity. In this view the formation of junctions may be considered a novel feature of the network of cosmic superstrings.  Different theoretical aspects of (p,q) strings construction were studied in 
\cite{Copeland:2003bj, Firouzjahi:2006vp, Jackson:2004zg, Copeland:2006if}
while the cosmological evolution of a network of strings with junctions have been investigated in \cite{Tye:2005fn}. 

The most important cosmological implications of cosmic strings are the gravitational ones which are
controlled by $G\mu$. Among them are the lensing effects. The geometry around a cosmic
string is locally flat but globally it produces a deficit angle in the plane perpendicular to 
the string \cite{Vilenkin:1981zs}.  An observer looking at an object behind the string may see two identical images located on opposite sides of the string. 

In this paper we study the lensing and CMB anisotropies due to an arbitrary configuration of straight cosmic strings forming
a stationary junction. We will provide the sum rule for the formation of multiple images. The effects of cosmic string wakes on structure formation is briefly studied.

\section{The setup}

We are interested in the metric of strings at a stationary junction. We assume the junction is at rest. With an appropriate boost, one can also consider the case of a stationary junction moving with a constant velocity.

The action of $N$ semi-infinite strings joined at a point is
\ba
S= -\sum_{i} \mu_{i} \int d\, t  \, d \, \sigma \sqrt{-|\gamma_{i} | }  \, ,
\ea
where $\mu_{i} = | \vec \mu_{i} |$  is the tension of the  i-th string 
(the vector pointing in direction of the string)
and  $\gamma_{i\, mn}$ is the metric induced on each string
\ba
\gamma_{i\,mn} = g_{\mu \nu}  \partial_{m}X_{i}^{\mu} \, \partial_{n} X_{i}^{\nu} \, .
\ea
Here ${m,n} =\{ t, \sigma \}$ are the coordinates along the string worldsheet, $X^{\mu}$ are the space-time coordinates and $g_{\mu \nu}$ is the space-time metric.
We shall choose $\sigma=0$ at the junction and increasing away 
from this point.
Furthermore, $X^{0}=t$ while $X^{1}_{i}=x_{i}, X^{2}_{i}=y_{i}$ and $ X^{3}_{i}=z_{i}$.

In order for the junction to be stationary, the vector sum of the tensions should
vanish at the junction: $\sum_{i} \vec \mu_{i}=0$. 
Suppose the junction is the origin of the coordinate system and 
the unit vector along the i-th string is denoted by $\vec n_{i}$. The conditions for the junction to be stationary are translated into
\ba
\label{station}
\sum_{i} \mu_{i} \, n_{i\, x} =
 \sum_{i} \mu_{i}\,  n_{i\, y} =
 \sum_{i} \mu_{i} \,  n_{i\, z}=0 \, .
\ea

To solve the Einstein equations, we need to find the energy-momentum tensor $T^{\mu \nu}$ for the string configuration, obtained by varying the
string action with respect to the metric
\ba
\label{deltaS}
\delta_{g} S= -\frac{1}{2} \sum_{i} \mu_{i} \int d\, t  \, d \, \sigma \sqrt{-\gamma_{i}} \,
\gamma_i^{mn} \partial_{m}X_{i}^{\mu} \, \partial_{n} X_{i}^{\nu} \delta g_{\mu \nu}
= -\frac{1}{2} \int d^{4}x \, T^{\mu \nu} \delta g_{\mu \nu} \, ,
\ea

For each string, one can choose $\sigma$ to represent the line element 
along the string
\ba
d\, \sigma^{2}= d x_{i}^{2} + d y_{i}^{2} + d z_{i}^{2} \equiv d l_{i}^{2}
\, ,
\ea
which implies
\ba
{\ga_{i}}_{00}=1   \quad , \quad {\ga_{i}}_{\sigma \sigma}=-1
\quad , \quad  {\ga_{i}}_{0 \sigma} = 0 \, .
\ea 
Using these in Eq. (\ref{deltaS}) one obtains 
\ba
\label{stress}
T^{\mu \nu}(\x)= \sum_{i} \,  \mu_{i} \,  \int d l_{i} \,    \gamma_i^{mn}\,  \partial_{m}X_{i}^{\mu} \, \partial_{n} X_{i}^{\nu}   \, \delta^{3} (\x - \x_{i} )\, ,
\ea
where here and in the following, the boldface letters represent the spatial parts of the four-vectors 
$X^{\mu}$.

We start from a flat background metric $\eta_{\mu \nu}$ = diag $(1,-1,-1,-1)$ . Up to linearized level, the metric is
\ba
g_{\mu \nu} = \eta_{\mu \nu} + h_{\mu \nu} \, ,
\ea
where $h_{\mu \nu}$ is the perturbation due to cosmic strings, $|h_{\mu \nu}| <<1$.

The linearized Einstein equations are
\ba
\label{Einstein}
R_{\mu \nu} =   8 \pi G \, (T_{\mu \nu} -\frac{1}{2} \eta_{\mu \nu} T)  \equiv  8 \pi G\,  S_{\mu \nu} \, ,
\ea
where $R_{\mu \nu}$ is the Ricci tensor and $T= T^{\mu}_{\mu}$ is the trace of the  energy-momentum tensor.  Furthermore, using (\ref{stress}) one obtains $S_{0\mu}=0$ while 
\ba
\label{Sab}
S_{ab}(\x)= \sum_{i} \mu_{i} \,  (\delta_{a b} - n_{i\, a} \, n_{i\, b} )  \int d l_{i}  \, \delta^{3} (\x - \x_{i} )\, ,
\ea
where $a, b$ are the spatial indices.

The Einstein equations as usual are subject to the choice of gauge. We use the harmonic gauge where 
\ba
\label{gauge}
{h^{\mu}_{\nu}}_{,\mu} - \frac{1}{2}  \, {h^{\mu}_{\mu}}_{,\nu} =0 \, ,
\ea
and
\ba
\label{Ricci}
R_{\mu \nu} = -\frac{1}{2} \,  \Box h_{\mu \nu} \, ,
\ea
where $\Box$ is the four-dimensional Laplacian. 

The general solution of (\ref{Einstein}) and (\ref{Ricci}) is (for example see \cite{Weinberg} )
\ba
\label{h}
h_{\mu \nu}( \x) =  - 4 G \, \int d^{3} \y  \, 
\frac{S_{\mu \nu} (t- | \y - \x | , \y ) }{ | \y - \x |  } \, .
\ea
The term $t- | \y - \x |$ stands for the retarded time. For our case of a static junction we are interested in the time-independent solution.

Using (\ref{Sab}) in (\ref{h}) yields $h_{0\mu}=0$ while 
\ba
h_{ab}(\x) = - 4 \, G \sum_{i} \mu_{i} \,  (\delta_{a b} - n_{i\, a} \, n_{i\, b} ) 
\int_{0}^{\infty}  \frac{d \l_{i}}{ | \x - \x _{i}|  } \, .
\ea
Knowing that along each string $\x_{i} = l_{i} \, \vec n_{i}$, the 
above integral can be calculated, and our final solution is
\ba
\label{h4}
  h_{ab} = 4 \, G \,   \sum_{i}  \,  \mu_{i} \, (\delta_{a b} - n_{i\, a} \, n_{i\, b} ) 
   \, \ln \left( \frac{r- {\vec r} \, . \, \vec n_{i}}{r_{0} }   \right) \, ,
\ea
where $r= |\x|$ is measured from the point of the junction and $r_{0}$ is a constant of integration.
One can directly check that the solution (\ref{h4}) satisfies the harmonic gauge (\ref{gauge}).

Similar to what happens in the case of a single infinite string, at points on each string where $\vec r= r\, \vec n_{i}$, the metric is singular. 
This is because we have started with delta function sources. In the
realistic situation when the strings have finite width, this singularity 
is smoothed out \cite{Gregory}.
On the other hand, at points on the opposite side of each string where $\vec r= -r\, \vec n_{i}$, the metric is non-singular.

Our solution in Eq. (\ref{h4}) is valid for any arbitrary configuration of 
straight cosmic strings in a stationary junction.
In order to understand its general applicability, let us
consider the case of a single infinite straight string, the case
originally considered by Vilenkin \cite{Vilenkin:1981zs}. This configuration 
in our formalism corresponds to two semi-infinite strings with equal tension extended back to back
\cite{Jackson} with $\vec n_{1} =-\vec n_{2}$. 
We may choose the strings to extended oppositely along the $z$-axis. 
We obtain $h_{zz} = h_{zx}=h_{zy}=h_{xy}=0$, while
\ba
 h_{xx} = h_{yy} &=&  4 \, G \, \mu \, \ln \left( \frac{ r^{2} - (r. \vec n_{1})^{2} }{r_{0}^{2}  }  \right) \nonumber\\
 &=&  8 \, G \, \mu \, \ln \left( \frac{r_{\perp}}{r_{0} }   \right) \, .
\ea
But $r_{\perp} $ is the normal distance to the string
from the point of the observer at position $\x$, and we obtain 
Vilenkin's solution \cite{Vilenkin:1981zs}. 

\section{The flatness of the geometry}

The geometry around an infinite cosmic string is flat away
from the string core. One may ask 
whether or not this is also true in our case of cosmic strings at a
static junction. At first sight, the metric given in (\ref{h4}) does not
seem to be flat. To address this question we need to calculate the components of the Riemann tensor. 
Since the time coordinate decouples from our solution, effectively we are dealing with
a three-dimensional spatial geometry. In three dimensions, both the Riemann tensor $R_{abcd}$ and the Ricci tensor $R_{ab}$ have 6 independent components. This indicates that the components of Riemann tensor can be expressed in terms of Ricci tensor. More explicitly
\ba
\label{Riemann}
R_{xzzy}=R_{xy} \quad  , \quad R_{xyyz}=R_{xz} \quad , \quad
R_{yxxz}=R_{yz} \quad , \quad 
R_{xyxy}=\frac{1}{2}(R_{zz}-R_{xx}-R_{yy}) \, ,
\ea   
while the remaining two components $R_{xzxz}$ and $R_{yzyz}$ are obtained by the appropriate
permutations of the $x, y$ and $z$ coordinates.

Our solution  is a vacuum solution with the strings as sources. It is clear from (\ref{Einstein})
that $R_{ab}=0$ away from the strings. Using this in (\ref{Riemann}) we can immediately conclude that
all components of the Riemann tensor vanish and the geometry given by the metric (\ref{h4}) is indeed flat. 

The fact that the junction is stationary is the crucial requirement for the 
flatness of the geometry. It is evident that for non-stationary junctions 
the space-time is curved. The extent of the departure from a flat geometry
is directly controlled by the extent of violation of the stationarity 
conditions. For example, using the metric (\ref{h4}) one can show that
\ba
\label{Riemann2}
R_{xyxy}= \frac{-2G}{r^{3}}\, \sum_{i} \mu_{i} \,
 ( n_{i\, x} \, x+  n_{i\, y}\, y- \, n_{i\, z}\, z  ) \, .
\ea
This is zero due to the force balance conditions (\ref{station}).

\section{The Propagation of Light}

One of the novel cosmological features of cosmic string is the lensing 
effect. The metric of a straight cosmic string is locally flat. But 
globally the geometry around the string has a deficit angle given by
$\Delta= 8 \pi G \mu$. This results in the formation of two identical images of an object located behind the string. When looking
at an object located behind a static junction of semi-infinite strings, 
one naturally expects multiple images to form.
 
The lensing by three co-planar strings forming a {\bf Y}-shaped junction 
was studied by Shlaer and Wyman \cite{Shlaer:2005ry}. An observer looking 
at an object, say a galaxy, located behind the plane of the strings will see three (identical) images; one image is the object itself and the other two are its lensing counterparts.

In the method used in \cite{Shlaer:2005ry}, one starts with three infinite
strings intersecting at the point of the junction. This will
produce two {\bf Y}-shaped junctions oriented oppositely at the junction. One can ``cut-and-paste'' one junction and keep the remaining one. This
method will correctly produce the lensing by the junction, as demonstrated
in \cite{Shlaer:2005ry}.

In this section we would like to consider the general case of an arbitrary
number of stationary cosmic strings forming a junction, co-planar or not, 
and study the resulting lensing phenomena in cosmology.


\begin{figure}[t] 
\vspace{-2cm}
   \centering
   \hspace{-1.8cm}
  \includegraphics[width=3.2in]{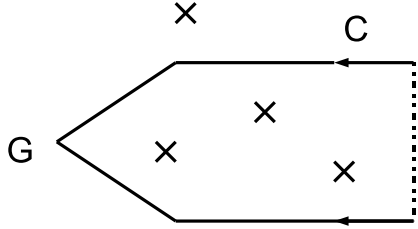} \hspace{1cm}
    \vspace{-5.5cm}
\caption{In this figure two parallel light rays emitted from infinity are deflected towards the point G. 
The contour C is made of the point G, the light rays and the line
connecting the two light rays distance at infinity, denoted by the
dashed line. The strings cross the plane spanned by the contour at the
points indicated by the cross signs. Only those strings which are enclosed
by the contour contribute in (\ref{v1}) and (\ref{mueff}).  }
\vspace{0.8cm}
\label{lensing1}
\label{lensing1}
\end{figure}


Suppose two parallel light rays are emitted from infinity towards the junction. Once the line connecting the light rays passes a string, the 
light rays are expected to be bent towards each-other. Suppose the light
rays stay co-planar and meet at a point, say G.  The difference in 
the velocity vectors at the point of intersection G can be obtained by the 
method of parallel transportation around a closed curved C. The curve C is composed of the point G, the two light rays
from infinity and the line connecting these rays at infinity.
For a schematic view see {\bf Fig. \ref{lensing1}}. The difference in the
velocity vectors at $G$ is \cite{MTW}
\ba
\label{v1}
\delta v^{\alpha } =- \frac{1}{2}  \int_{S} R^{\alpha}_{ \beta \gamma \lambda}\, v^{\beta} \, dx^\gamma \wedge dx^\lambda \, ,
\ea
where the integration is over the closed surface S bounded by C. 

Of course, in the empty regions away from strings the Riemann tensor 
vanishes as we shown before and the above integral is zero, as expected. 
However, when the surface intersects strings
the Riemann tensor provides delta-function contributions and the integral 
does not vanish. This  demonstrates that the relative change of the velocity vectors and consequently the lensing effects are directly controlled by the number and the orientation of the cosmic strings which
intersect the surface S. In the spirit, the method is analogous to the residue theorem for the integration of an analytical function in the complex plane.

Without loss of generality, we may suppose that the light rays are emitted
along the negative z-direction in the y-z plane. Using Eq. (\ref{Sab}) one obtains
\ba
\label{newS}
S_{a b} = \sum_i \frac{\mu_i}{n_{i\, x}} \, 
(\delta_{ab} - n_{i \, a} n_{i\,b} ) \,  \delta (y-y_i) \,  \delta(z-z_i) 
\, ,
\ea
where $y_i$ and $z_i$ are the coordinates of the point of intersection of 
the i-th string with the y-z plane. The factor $n_{{i\, x}}$ in the 
denominators originates from replacing $d l_{i}$ along the string by $dx$ 
via $dx= n_{i\, x}  \, d l_{i}$.

Using this in Eq. (\ref{v1}) one obtains
\ba
\label{vx}
\delta v^x&=& \int dy \, dz \, R_{xzzy} = \int dy \, dz \, R_{xy}
= 8 \pi \, G \int dy \, dz \, S_{xy} \nonumber\\
&=& - 8 \pi \, G \sum_i \mu_i\, n_{i\, y} \, ,
\ea
where the sum is over those strings which intersect the plane of the light rays (the y-z plane here) and are enclosed by the surface S bounded
by the light rays.

Similarly, for the change of velocity in the $y$-direction one obtains
\ba
\label{vy}
\delta v^y&=& \int dy \, dz \, R_{yzzy}= 
\frac{1}{2} \int dy \, dz \, ( R_{yy} + R_{zz} - R_{xx} ) \nonumber\\
&=& 8 \pi \, G \sum_i \mu_i\, n_{i\, x} \, ,
\ea
where to obtain the last equation, the identity 
$n_{i \,x }^2+n_{i \,y }^2+n_{i \,x }^2=1 $ has been used.

Finally, from (\ref{v1}) one can easily see that $\delta v_z=0$, which implies that there is no change of velocity in the direction tangential to the initial light rays.

Combining these results, one obtains the following coordinate independent
representation of the change in velocity
\ba
\label{deltav}
\delta \vec v= -8 \pi \, G \, \vec k \times \sum_i \vec \mu_i \, ,
\ea
where the unit vector $\vec k$ represents the direction of the light rays
at infinity (line of sight).

The angle between the two light rays at the point of intersection  is 
\ba
\label{Delta2}
\Delta = | \delta \vec v |= 8 \pi \, G \,  \mu_{eff} \, ,
\ea
where
\ba
\label{mueff}
\mu_{eff}= | \vec k \times \sum_i    \vec \mu_{i}  |  \, .
\ea
As explained before, the sum is over those strings which intersect
the plane S formed by the light rays up to their point of intersection.
For the case of a single string with tension $\mu$ enclosed between the light rays, $\mu_{eff}=\mu$ and we obtain the standard result for the deficit angle.

\begin{figure}[t] 
\vspace{-2cm}
   \centering
  \includegraphics[width=3in]{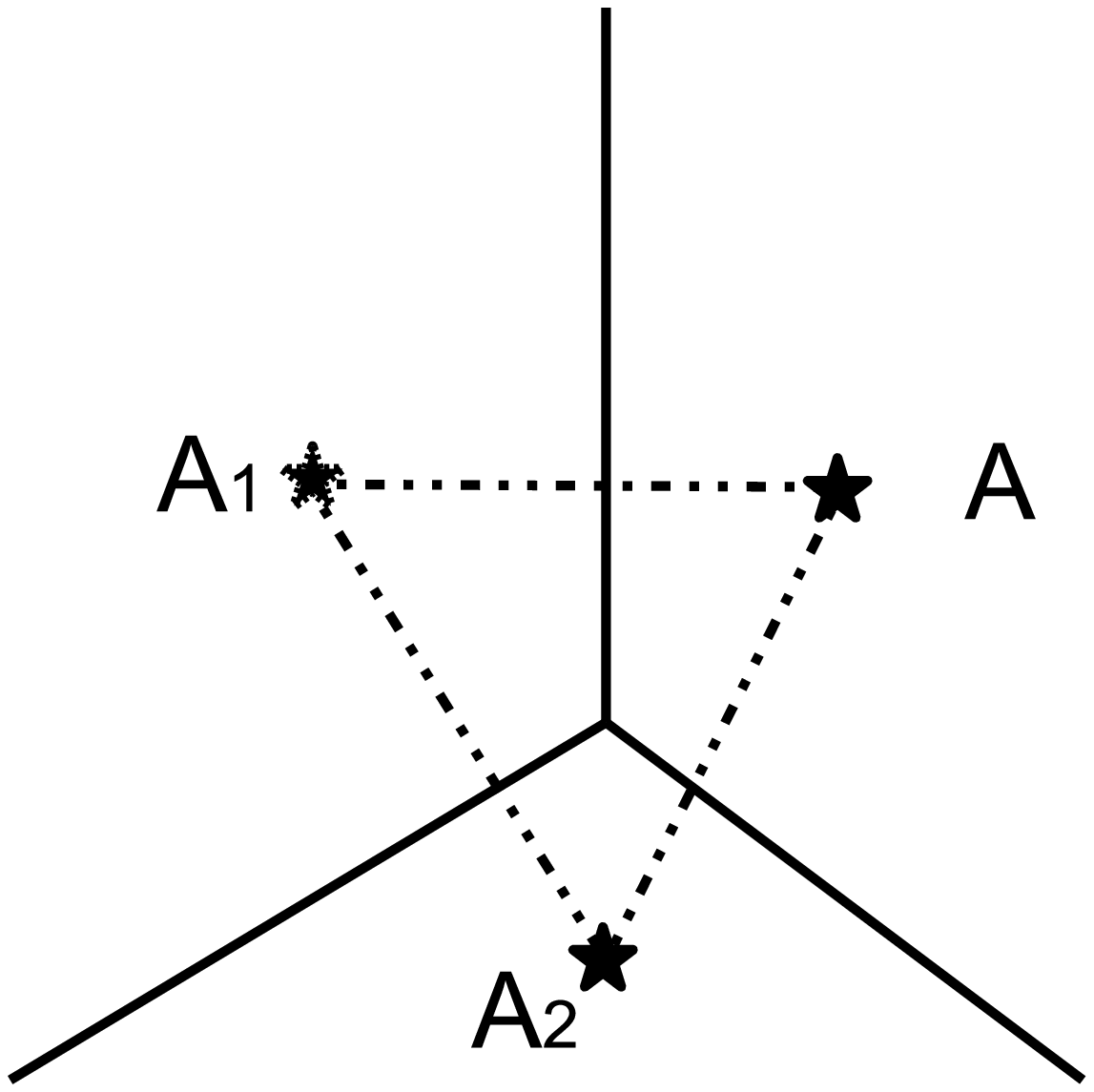} \hspace{-1cm} \vspace{0.5cm}
    \includegraphics[width=3.2in]{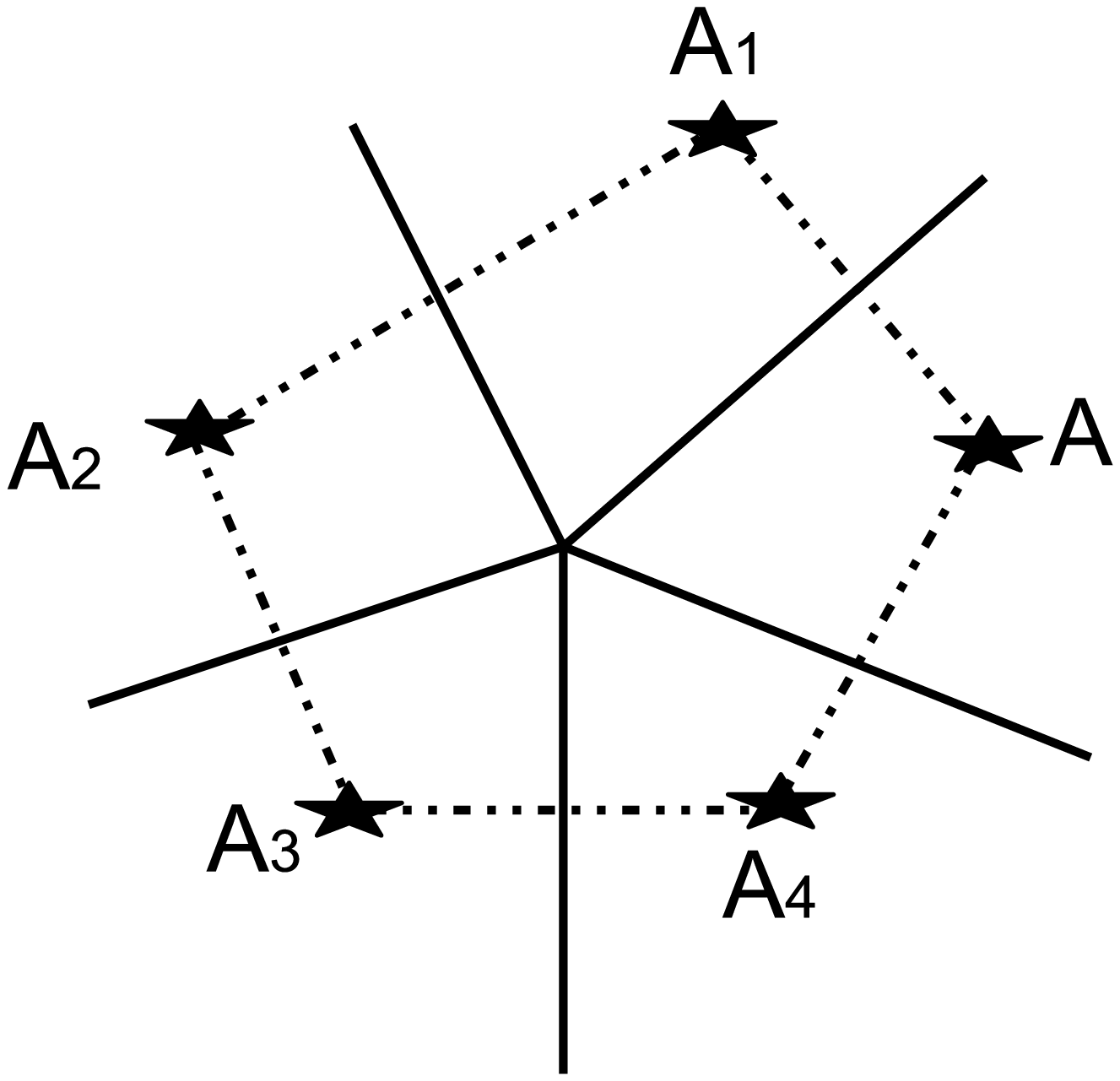} \hspace{-1cm}
    \vspace{-3.5cm}
\caption{In this figures the multiple lensings by $N$ co-planar strings at
 a junction is sketched. The object A and its lensing counterparts 
$A_{i}, i=1..N-1$, form a closed loop.  Each image is the source for 
the nearby images via the enclosed string.  For example, in the 
three-string junction in the left figure, the object A is the source for 
the image $A_{1}$ via the string $\mu_{1}$, $A_{1}$ is the source for the 
image $A_{2}$ via the string $\mu_{2}$ and $A_{2}$ is the source for the 
object A via the string $\mu_{3}$.}
\vspace{0.5cm}
\label{YNjunction}
\end{figure}

Now let us try to apply Eq. (\ref{Delta2}) to some examples.
The first interesting example is the lensing by a {\bf Y}-shaped junction 
studied in \cite{Shlaer:2005ry}. The strings are in the x-y plane and the 
light rays are emitted along the z-direction.
A schematic view of this situation is presented in
{\bf Fig. \ref{YNjunction} }. Starting with the object located at point 
$A$ and moving in counterclockwise direction, we enclose the string with
tension $\mu_1$ and  $\mu_{eff}=\mu_1$. Consequently, there is an image at 
point $A_1$ separated from $A$ by the deficit angle 
$8 \pi \, G \,  \mu_{1}$. Continuing further counterclockwise, we enclose 
a second string (with tension $\mu_2$) and 
$\mu_{eff}= |\vec \mu_1 + \vec \mu_2|$. Because of the force balance condition this is the same as $\mu_3$. So the second image is at the 
point $A_2$ which is separated from the object $A$ by a deficit angle 
$8 \pi \, G \,  \mu_{3}$. Finally, continuing further to enclose the third
string, the effective tension vanishes
due to force balance condition $\vec{\mu_1} + \vec{\mu_2} + \vec{\mu_3}=0$
and there is no other image. This means that the points $A, A_1$ and $A_2$
form a triangle. This has an interesting geometrical interpretation: 
Each image acts as a source for the other two images. The object $A$ is 
the source for $A_{1}$ through the string $\mu_{1}$, the image $A_{1}$ 
acts as source for $A_{2}$ via the string $\mu_{2}$ and the image $A_{2}$
is the source for $A$ via the string $\mu_{3}$.
This is in exact agreement with the prescription provided for 
a {\bf Y}-shaped junction in \cite{Shlaer:2005ry}. 

In this example the strings are between the observer and the object and 
the plane of the strings is perpendicular to the line of sight. 
Whether or not the observer actually sees an image depends on the angular
distances between each string and the object. If a distance is larger 
than the deficit angle, then the corresponding image is not
observable. 

One can generalize the example of the {\bf Y}-shaped junction
to the case of $N$ co-planar cosmic strings at junction.  As before, 
the plane of the strings is between the object and the observer and is
perpendicular to the line of sight. Following the same steps as above, we
obtain $N-1$ images plus the object itself. This set of $N$ points forms a
closed loop due to the force balance condition. 
For each image, the effective tension is the magnitude of the vector sum 
of all strings enclosed between the object and the corresponding image.
Geometrically, as before, this means that each image is the source for the
nearby images via the enclosed string. Starting with the object $A$, the chain of object $\rightarrow$ image is given by $A \rightarrow A_{1}\rightarrow A_{2} \rightarrow .. \rightarrow A_{N-1} \rightarrow A$.
The  schematic view of this case is presented in {\bf Fig. \ref{YNjunction}}.                                                                                                         


Another interesting case is when the light rays are parallel to the plane 
of the co-planar strings. Now the strings are located in the y-z plane and 
two light rays are emitted along the z-direction. If the light rays are on 
the same side of the plane, then no string is enclosed by them and the 
light rays will stay parallel. Now suppose the light rays are emitted from
opposite sides of the plane. To simplify the analysis suppose the light
rays are in the x-z plane so they are emitted 
at $(x_{1}, y_{1},  \infty)$ and $(x_{2}, y_{1}, \infty)$  with 
$x_{1} x_{2} <0$. 
We already know that $\delta v_{z}=0$, i.e. there is no change of velocity 
in the direction parallel to the light rays.
Also, from Eq. (\ref{vy}) one obtains $\delta v_{y}=0$ because the strings 
have no $n_{x}$ components. On the other hand, from Eq. (\ref{vx}) one 
obtains 
\ba
\label{parallel}
|\delta v_{x}| =    8 \pi \, G | \sum_i \mu_i\, n_{i\, y} | \, ,
\ea
where the sum is over the strings in the upper part of the y-z plane. 
From the force balance condition, this
is also equivalent to the sum over the strings in the lower part of the 
y-z plane.
Interestingly enough, we see that the light rays which start out parallel
to the plane of strings are now deflected towards the plane. 


\section{CMB anisotropies and cosmic String wakes}

The effects of moving string on CMB anisotropy were studied by Kaiser 
and Stebbins \cite{Kaiser:1984iv}. These authors showed that as a
consequences of the conical structure of space perpendicular to the
string, a moving string will produce a line discontinuity in temperature
anisotropy maps, with the amplitude of the discontinuity given
by 
\ba
{{\delta T} \over T} \, = \, G \mu \gamma(v) v \, , 
\ea
where $v$ is he string velocity perpendicular to our line of
sight towards the string, and $\gamma(v)$ is the Lorentz factor.
The effect is due to the relative Doppler shift in the photons passing
on the two sides of the string. The Kaiser-Stebbins effect is a
key distinctive observational signature for strings. Methods to
search for this signature have recently been discussed in
\cite{ABB,KSobs} (see also \cite{Richhild} for older work).

One may naturally ask how this can be generalized to the case of strings 
at a junction when the junction is moving with a constant velocity $v$. 
It is understood that the junction and the strings attached to it
move as a solid object and the no force condition (\ref{station}) still
holds.

We are interested in change in the observed frequency between two 
parallel light rays (with the same initial frequency) in the presence 
of the moving junction. As before, we take the light rays to be moving 
in the y-z plane, and initially along the z-direction.
In terms of Eq. (\ref{v1}) instead of $v^{\mu}$ we use the momentum 
four-vector
$p^\mu= ( E, {\bf p})$, where $E= | {\bf P}| = \hbar \omega$. This yields
\ba
\label{v0}
\delta \omega \, =  \, \omega  \int R_{0zzy} \, dy \, dz  \, .
\ea
Unlike in the case of a static junction, the Riemann tensor has non-zero
components like $R_{tabc}$ and $R_{tatb}$ due to the string motion. 
In 4-D space-time the Riemann tensor has more components than the Ricci
tensor does. However, we can go to the junction's rest frame where the
space-time is static and our results from the previous section can be used 
readily. Denote the coordinate system where the junction is static by 
${x^{\mu'}}$ while the coordinate system used by the observer and the light 
sources is represented by ${x^{\mu}}$. 
We have
\ba
R_{0 zzy } = \frac{\partial \,  x^{\alpha'}}{\partial  \, t} \frac{\partial \, x^{\beta'}}{\partial \, z} 
\frac{\partial  \, x^{\mu'}}{\partial  \, z}  \frac{\partial \, x^{\nu'}}{\partial \,  y} R_{\alpha' \beta' \mu'\nu'} \, ,
\ea
where now $ R_{\alpha' \beta' \mu'\nu'}$ has no time-like indices.

In general the junction velocity vector $\vec v$ can have components in x,
y and z-directions. Suppose the junction is moving along the x-direction.
Then one obtains
\ba
\frac{\delta \omega}{\omega} &=& \left| \int d\, y' \, d \, z' \, \frac{\partial \, x'}{ \partial \, t} R_{x'z'z'y'} \right| \nonumber\\
&=& 8 \pi G \gamma \, v \left|  \sum_{i } \mu_{i}  \, n_{i\, y}  \right| \, .
\ea
To obtain the final line, the same procedure as in Eq. (\ref{vx}) has
been used. As before, in the above the sum is over the strings which are 
enclosed by the light rays. 

Similarly, in the case when the junction is moving along the y-direction one obtains
\ba
\frac{\delta \omega}{\omega} &=& \left| \int d\, y' \, d \, z' \, \frac{\partial \, y'}{ \partial \, t} R_{y'z'z'y'} \right| \nonumber\\
&=& 8 \pi G \gamma \, v \, \left| \sum_{i } \mu_{i}  \, n_{i\, x} \right | \,  ,
\ea
where to obtain the final result Eq. (\ref{vy}) was used.

Finally, if the junction is moving along the z-direction parallel to the
light arrays one can show that $\delta \omega$ vanishes.

Combining all these results, one can show that
\ba
\label{cmb}
\frac{\delta \omega}{\omega} &=& 8 \pi G \,  \gamma  \,  
\left| \,  \vec v \,  . \, ( \vec k \times \sum_{i }\vec \mu_{i} ) \right| \, ,
\ea
where as before $\vec k$ represents the line of sight and the sum is over
the strings which are enclosed by the light rays. Interestingly enough,
this formula has the same functional form
as that of the a single string \cite{Vachaspati:1986gf}.

The implications of Eq. (\ref{cmb}) for the CMB anisotropies are
parallel to those of Eq. (\ref{Delta2}) for lensing. Each string at the 
junction produces its own CMB anisotropies. The change
in the CMB temperature across the i-th string is given by 
\ba
{{\delta T} \over T} \, = \,
8 \pi G \, \gamma \,  |   \vec v  . (  \vec \mu_{i} \times  \vec k) | \,. 
\ea
Interestingly enough, the temperature
anisotropy is different for different legs of the strings at the junction, depending on the tensions and orientations of the strings.

We conclude that the distinctive signature of strings with junctions for
CMB anisotropies is the possibility of line discontinuities joined
at a point for an observer looking at the surface of last scattering. 
The details of this picture depends on the orientations of the
strings and the direction of its motion of the junction with respect to 
the line of sight. In recent work \cite{ABB}, the Canny 
algorithm \cite{Canny}, an edge detection
algorithm, was shown to yield a sensitive statistic with which
CMB line discontinuities can be identified in the sky. Since
string junctions are a prediction of cosmic superstrings (simple
gauge theory strings do not admit junctions), it would be interesting
to develop modified edge detection algorithms which can differentiate
between strings with and without junctions in CMB anisotropy maps.
Work on this issue is in progress.


The lensing produced by the cosmic string deficit angle also produces
distinctive signatures for structure formation, namely ``wakes"
\cite{wakes}. For the rest of  this section, we will briefly discuss the 
implications of string junctions for cosmic string wakes.
Thus, we consider the accretion of matter by a moving junction.
 
Consider two massive non-relativistic objects at rest in the
frame of the cosmic microwave background in the presence of a 
moving junction. 
As in the previous section, it is convenient to
go to the rest frame of the junction. In this frame, the two massive objects are moving
towards the junction. When the line connecting the objects passes any 
string, the objects are attracted towards each other behind the cosmic 
string. This can provide a mechanism for structure formation by string 
wakes \cite{wakes,Stebbins:1987cy,Sornborger}. 

In the case of a single infinite string, the lensing of matter behind
a moving string leads to a region of twice the background density
in the wake of the moving string. The opening angle of the wake
is given by the deficit angle. The question we wish to address in
this section is how the wake structure generalizes to strings with
junctions.

The results obtained in Section ({\bf IV}) for the bending of two light
rays can be used for massive objects too. The only change is the addition
of $\gamma v$ factor due to the change of frame. The relative velocity
between the massive objects at the intersection is 
\ba
\label{wake1}
\delta \vec v &=& -8 \pi G \,  \gamma  \,  v
 \, \vec k \times \sum_{i }\vec \mu_{i}  \, .
\ea
On the other hand, the angle between the particles trajectories is 
$\Delta_m= | \delta \vec v | /v$, and one obtains
$\Delta_m = 8 \pi G \,  \gamma  \, 
|\,  \vec k \times \sum_{i }\vec \mu_{i}  \, | \, .$
This means that around each leg of the string junction there is a 
wedge-shaped wake with opening angle $\Delta_m$. In each of these 
regions, the matter density is doubled, i.e.
$\delta \rho/\rho=2$. 

Since the cosmic strings cannot be the dominant source of structure
formation, and since string wakes undergo non-trivial non-linear
evolution, it will presumably be more difficult to find distinctive
signatures for string junctions in large-scale structure surveys
than in CMB anisotropy maps. In principle, topological statistics
such as Minkowski functionals \cite{Minkowski} have the power to
find such signatures \cite{topology}.


\section{Conclusions}

Working at the level of linearized gravity, we have derived the
metric of a static string junction, with an arbitrary number of
strings joining. We have shown that away from the world sheet of
the strings, the metric is flat. Thus, the geometry generalizes
that for a single straight infinitely long string. Each string
segment produces a deficit angle, and thus deflects both light
and matter. We have derived the lensing of light by a string
junction, discussed the CMB anisotropies produced by a string
junction which is in uniform motion relative to the frame of
the microwave background, and commented on the formation of
string wakes.

We have identified a junction of line discontinuities in CMB
anisotropy maps as a distinctive signature of strings with
junctions, and proposed that one could look for these signatures
by a generalized edge detection algorithm. Finding positive
evidence for such string segments would provide a boost for
superstring theory, since it is in the context of superstring
theory that the existence of strings with junctions first came
to prominence. Strings with junctions are quite generic in the
context of superstring theory, but they do not appear in simple
gauge field theory models of cosmic strings.



\begin{acknowledgments}{ We would like to thank K. Dasgupta, O. Saremi and H. Tye 
 for useful discussions. This work is supported by NSERC, the Canada Research Chair Program and an FQRNT Team Grant. }\\
\end{acknowledgments}

\end{document}